\documentstyle[epsf]{cupconf}

\newcommand{\kms}{km~s$^{-1}$ }

\newcommand{\ApJ}{ApJ}
\newcommand{\AaA}{A\&A}
\newcommand{\JFM}{J. Fluid Mech.}
\newcommand{\ARFM}{Annual Review of Fluid Mechanics}

\newcommand{\bfi}{\begin{figure}}
\newcommand{\efi}{\end{figure}}

\title[Turbulence in Ursa Major]{Turbulence in the Ursa Major cirrus cloud}

\author[M.-A. Miville-Desch\^enes {\it et al.\/}]%
{M\ls A\ls R\ls C\ls -\ls A\ls N\ls T\ls O\ls I\ls N\ls E\ns M\ls I\ls 
V\ls I\ls L\ls L\ls E\ls -\ls D\ls E\ls S\ls C\ls H\ls \^E\ls N\ls E\ls 
S$^{1,2}$,\\
G\ls .\ns
J\ls O\ls N\ls C\ls A\ls S$^{2}$
\and\ns E\ls .\ns
F\ls A\ls L\ls G\ls A\ls R\ls O\ls N\ls E$^{3}$}

\affiliation{$^1$Institut d'Astrophysique Spatiale, B\^at. 121, Universit\'e Paris XI, F-91405 Orsay, 
France \\[\affilskip]
$^2$ D\'epartement de Physique, Universit\'e Laval and Observatoire du mont M\'egantic, Qu\'ebec, Qu\'ebec, Canada, G1K 7P4\\[\affilskip]
$^3$ Radioastronomie, Ecole Normale Sup\'erieure, 24 rue Lhomond, 75005, Paris, France}

\begin{document}

\maketitle

\begin{abstract}
High resolution 21 cm observations of the Ursa Major cirrus revealed highly filamentary 
structures down to the 0.03 pc resolution. These filaments, still present in
the line centroid map, show multi-Gaussian components and seem to be associated with
high vorticity regions. Probability density functions of line centroid increments and structure
functions were computed on the line centroid field, providing strong evidences for the presence
of turbulence in the atomic gas.
\end{abstract}

\firstsection 

\section{Introduction}

Many statistical studies of the density and velocity structure of dense interstellar matter
have been done on molecular clouds where turbulence is seen as
a significant support against gravitational collapse that leads to star formation. 
Less attention has been devoted to turbulence in the Galactic atomic gas (HI).
The cold atomic component ($T \sim 100$ K, $n \sim 100$ cm$^{-3}$), alike molecular gas, is
characterized by multiscale self-similar structures and non-thermal linewidths. 

A detailed and quantitative study of the turbulence and kinematics of HI clouds has never been done. 
Here we present a preliminary analysis of this kind based on high resolution 21 cm 
observations of an HI cloud located in the Ursa Major constellation. 
To characterize the turbulent state of the atomic gas, a
statistical analysis of the line centroid field has been done. We have computed probability density 
functions of line centroid increments and structure functions.

\section{HI Observations}

The Ursa Major cirrus ($\alpha(2000) = 9h36m,\; \delta(2000) = 70^\circ20'$) has been observed
with the Penticton interferometer. Half of these data has already been published by 
Joncas et al. (1992). 
A total of 110500 spectra were taken on a 
$3.75^\circ\times2.58^\circ$ field (1' resolution) 
with a spectral resolution of $\Delta v = 0.412$ km s$^{-1}$.
This cirrus is part of a large loop of gas, possibly blown by stellar winds and/or supernovae explosions, 
known as the North Celestial Loop. A distance of 100 pc is adopted (\cite{de_vries87}).

\bfi
  \epsfbox{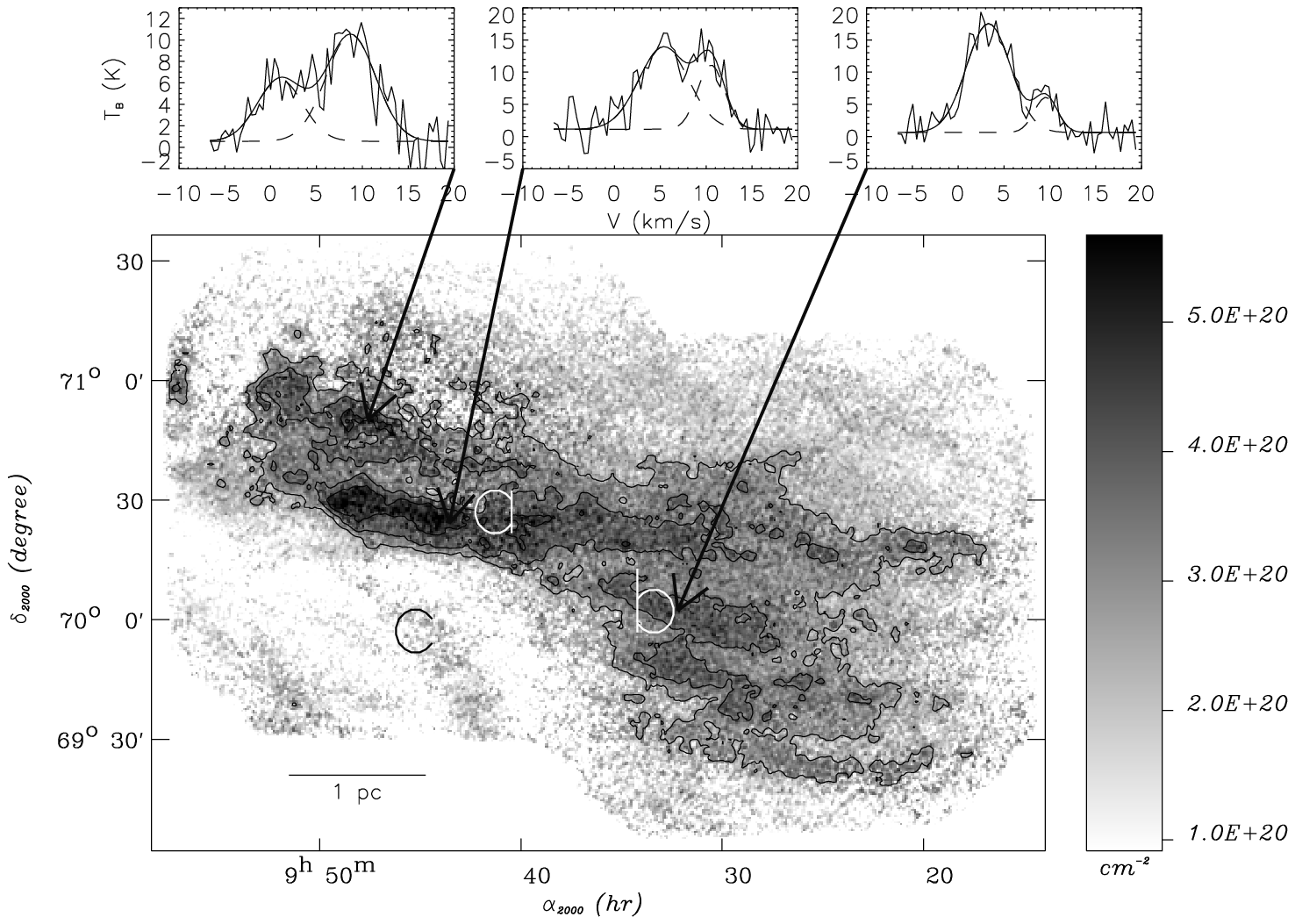}
  \caption{Integrated 21 cm intensity map expressed into HI column densities by the relation
	$N_{HI} = 1.8 \times 10^{18} \int T_B \, dv$} 
  \epsfbox{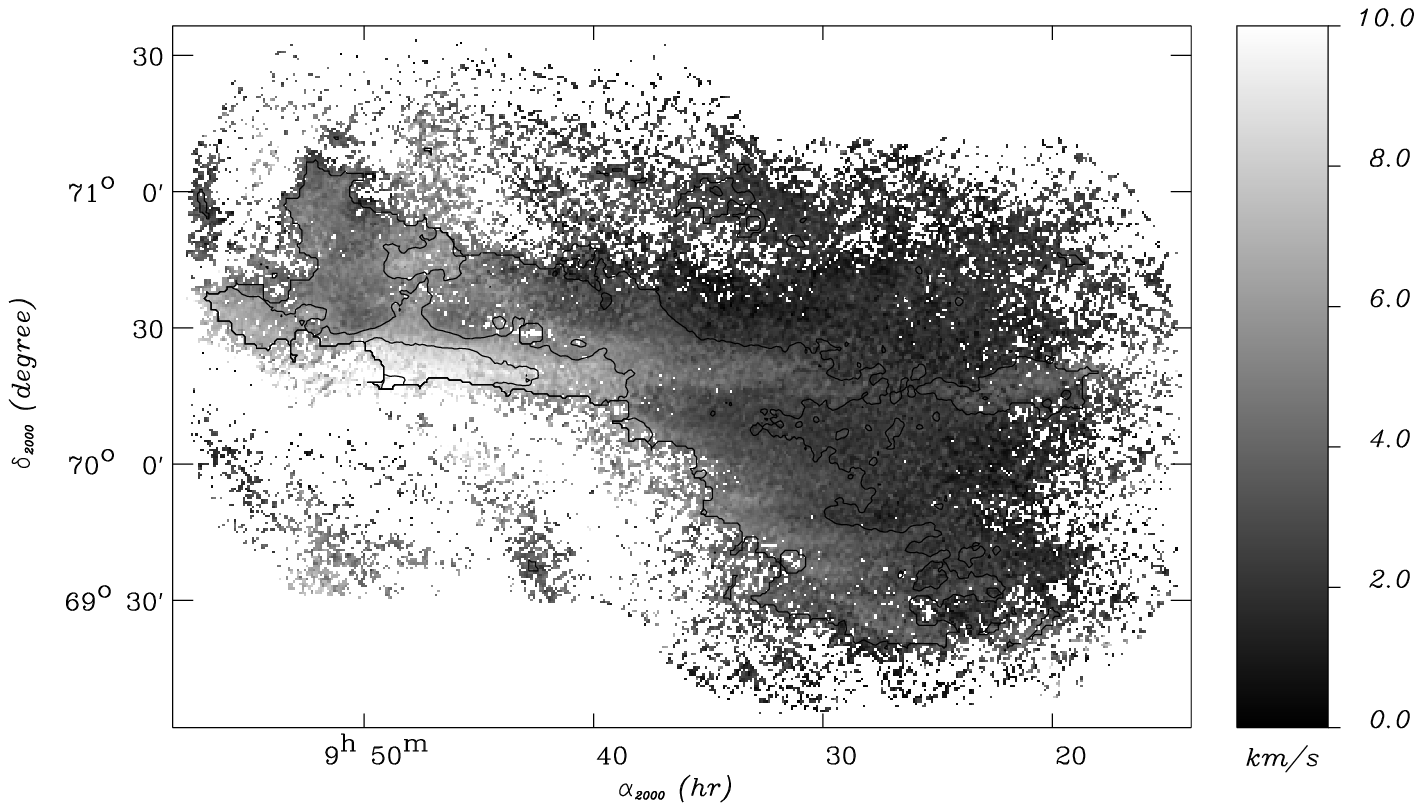}
  \caption{Map of line centroids for spectra with a signal-to-noise ratio greater than 4.} 
\efi

The integrated intensity map, proportional to the column density, is shown in figure~1; 
many filamentary structures are apparent. The filaments
stand out in individual channel maps and most of them can be followed over several channels. 
The strongest filament ({\bf a} on figure~1) crosses most of the field in an east-west direction.
There are also two fainter filamentary structures ({\bf b} and {\bf c}) that seem to merge
(in spatial and velocity space) in the brighter part of filament {\bf a}. 
A closer look at these filaments in individual channel maps reveals smaller elongated 
substructures, parallel to the main axis.

We performed a multi-Gaussian fit on each spectrum and found that most spectra
are fitted by one or a combination of two Gaussians of typical velocity dispersion $\sigma=2.8$ \kms.
The smallest velocity dispersion measured ($\sigma\sim 1.3$ \kms) is located in 
the northern part of filament {\bf c}. If this width were purely thermal it would correspond
to a kinetic temperature of 200 K. 
The spectra along filament {\bf a} are composed of two Gaussian components (see central spectrum
on figure~1). These two components are separated by $\Delta v=4.0$ \kms all along the filament
except in the eastern end where they merge. At this position we observe a 
clear velocity gradient (10.0 \kms per pc) across filament {\bf a}.
These two observational facts may be interpreted
as a rotation in the plane of the sky, along the major axis of the filament.
Assuming the filament is a cylinder, its radius is $R=0.2$ pc. Then, the turnover 
time ($\tau_t = 2R/\Delta v$), calculated from both behaviours (component separation and 
velocity gradient) is $\tau_t= 1\times 10^5$ year.

\section{Statistical analysis of the line centroid map}

Figure 2 displays the map of the line centroid ($v_0 = \int v\, T_B(v)\, dv/\int T_B(v)\, dv$).
On large scales, the line centroid field is characterized by a global east-west velocity gradient. 
On small scales, we note a significant difference between the eastern part 
(particularly near $\alpha=9h47m,\; \delta=71^\circ15'$) which shows velocity fluctuations on
scales of a few arcminutes and the western part which is very smooth at those scales. 
The brighter part of filament {\bf a} as well as filament {\bf b} and {\bf c} clearly 
stand out as coherent structures in the centroid~map.

\subsection{Probability Density Function of velocity increments} 

\bfi
  \epsfbox{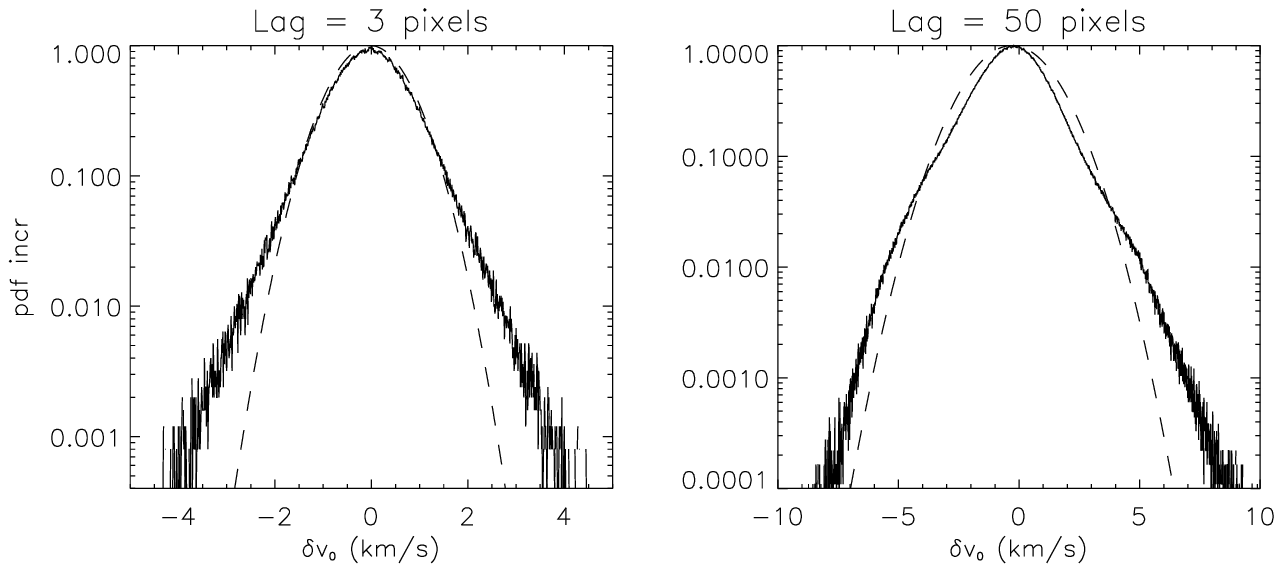}
  \caption{Normalized Probability Density Function of velocity increments (lag of 3 and 50 pixels)
	computed from the line centroid field of figure~2.} 
\efi

Probability density functions (PDF) of velocity increments and derivatives found
in experiments and numerical simulations of turbulent flows usually show clear non-Gaussian
wings. This non-Gaussian behaviour is attributed in part to coherent vortex structures
(\cite{sreenivasan97}).
In astrophysical context, non-Gaussian wings were observed by \cite{falgarone90} in molecular
line profiles of selected molecular clouds.
Recently, \cite{lis96} showed that PDFs of line centroid velocities 
{\it increments} (i.e. PDF of $\delta v_0 = v_0(r)-v_0(r+\delta r)$)
computed from numerical simulations of compressible turbulence at high Reynolds number 
have non-Gaussian wings.

Using the line centroid map shown in figure~2 we computed PDF of velocity increments
for 5 and 50 pixels lags.
As shown in figure 3, non-Gaussian wings are clearly present in the Ursa Major cirrus. 
Furthermore, exponential wings are stronger at small lag.
This is a behaviour typical of intermittency also observed in three dimensional simulations
of turbulence and in laboratory experiments (\cite{sreenivasan97}).
The points
that populate the non-Gaussian wings are essentially located along filaments {\bf a}, {\bf b}
and {\bf c}, and in the north-eastern part of the field. In their analysis 
\cite{lis96} showed a clear spatial correlation 
between the non-Gaussian events and regions of high vorticity.

\subsection{Structure Functions}

Structure functions (SF) of various orders ($<\Delta u^n> = \frac{1}{N}\sum \delta v_0^n$)
are extensively used to deduce physical 
parameters of turbulent flows (\cite{anselmet84}). Second and third order SF were computed
from the line centroid map. The large number of velocity points used here 
allows us to separate spatially our sample in two equal parts (east and west). 
We found a significant difference between the eastern and western SFs.

In the inertial sub-range of a Kolmogorov type turbulence, 
the exponent of the second and third order SF should be respectively 2/3 and 1.
Here, as seen in figure~4, the second order SFs  
are very smooth and show various exponents (from $n_2=0.25$ to $n_2=0.89$) on different scales 
and locations. The second order SFs drop near 1.5 pc; this may be caused by the poor 
statistics we have at large scale.
The third order SFs are also very smooth. The exponents range from $n_3=0.96$ to
$n_3=2.20$.

From 0.4 pc to 1.5 pc, the eastern part of the line centroid field 
is characterized by slopes of $n_2=0.77$ and 
$n_3 = 0.96$, and seems to meet the Kolmogorov type description for which the third order SF 
has an exact solution:
$$ <\Delta u_r^3> = -\frac{4}{5}r<\epsilon>$$
where $<\epsilon>$ is the mean energy transfer rate in the inertial range. 
This rate can be computed for the eastern part of the field: $<\epsilon> = 1\times10^{-2}$
erg s$^{-1}$ g$^{-1}$. On a small lag range (from 0.4 pc to 0.8 pc) we also
observed an exponent of 1 in the western third order SF, giving
$<\epsilon> = 0.8\times10^{-4}$ which is significantly smaller than the eastern value. 

\bfi
\hspace{-1.7cm}
  \epsfbox{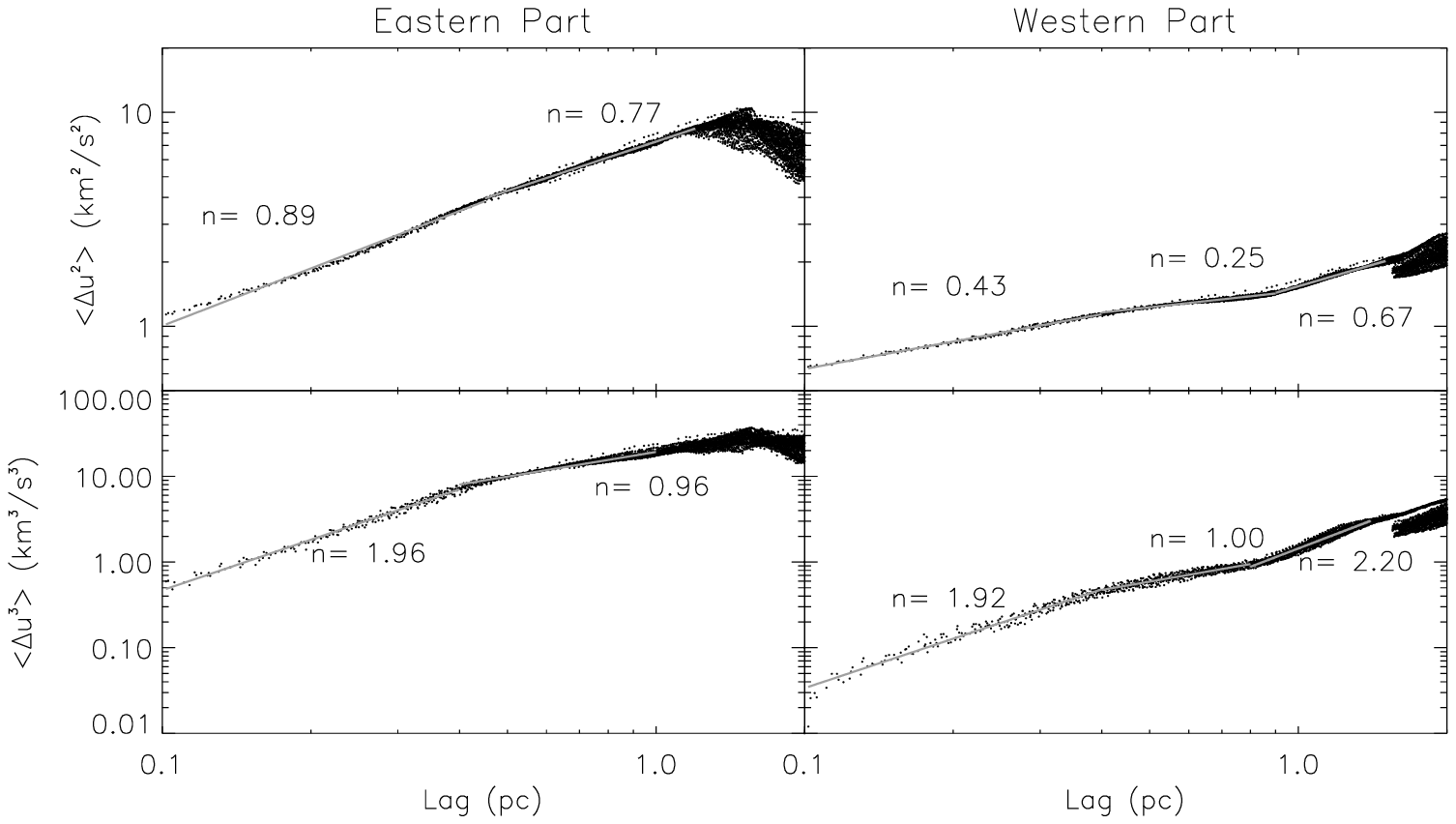}
  \caption{Second and third order structure function of the line centroid field. The structure functions
	were computed on the west and east side of the field}
\efi

\section{Discussion}

The previous statistical analysis supports the presence of turbulence in the Ursa Major cirrus.
The various observational results shown here
(the presence of clearly non-Gaussian wings in PDFs of line centroid velocity increments and
the slopes of the structure functions)
are recognized as signatures of turbulence in terrestrial experiments or numerical simulations.
On the other hand, astrophysical 
observations lack three dimensions in phase space; any detailed comparison with numerical simulations or 
laboratory results must be done with great caution. Here the 3D geometry of the cloud is 
unknown, it is thus premature to interpret further the various SFs' exponents. 

Numerical simulations and laboratory experiments of turbulent flows are also characterized by
filamentary vortical structures and dissipative sheets around them \linebreak
(\cite{sreenivasan97}). These highly localized dissipative regions 
may play an important role in the thermodynamics and chemical evolution of the 
interstellar medium (\cite{falgarone95}). Identifying such regions would be of 
considerable astrophysical interest.
The present study of high resolution 21 cm observations reveals filamentary
HI structures characterized by doubled-peak spectra and minor axis velocity gradient. These
features may be interpreted as a rotation around the major axis of the filament.
Furthermore, the main filaments observed here are associated with high vorticity zones
traced by the PDF of velocity increments. 

The filaments we have observed in the Ursa cloud cannot be the
coherent vortices which develop at scales close to the dissipation scale of
turbulence because the linear resolution of our observations (0.03 pc) is more
than 100 times larger than the dissipation scale in diffuse clouds
(see Falgarone, these proceedings). Instead, we may be observing a larger
structure, made of very small vortices, braided together, as the result of their
interaction and merging. This merging of small vortices into larger vortices
has been followed in the numerical simulations of Vincent and Meneguzzi (1994)
and Porter et al. (1994). This would explain why the velocity pattern is more complex 
than a simple rotation.

\end{document}